\documentclass[runningheads]{llncs}
\usepackage{graphicx}
\usepackage{tikz}
\usepackage{comment}
\usepackage{amsmath,amssymb} % define this before the line numbering.
\usepackage{color}
\usepackage{multirow}
\usepackage{bbm}
\usepackage{makecell}
\usepackage{subfigure}
\usepackage{graphicx}

\usepackage[accsupp]{axessibility}  % Improves PDF readability for those with disabilities.

\begin{document}
% \linenumbers
\pagestyle{headings}
\mainmatter
\def\ECCVSubNumber{16}  % Insert your submission number here

\title{Boosting COVID-19 Severity Detection with Infection-aware Contrastive Mixup Classification} % Replace with your title

\begin{comment}
\titlerunning{ECCV-22 submission ID \ECCVSubNumber} 
\authorrunning{ECCV-22 submission ID \ECCVSubNumber} 
\author{Anonymous ECCV submission}
\institute{Paper ID \ECCVSubNumber}
\end{comment}
%******************

% \begin{comment}
\titlerunning{Infection-aware Contrastive Mixup Classification}
\author{Junlin Hou\inst{1} \and Jilan Xu\inst{1} \and Nan Zhang\inst{2} \and Yuejie Zhang\inst{1}* \and \\ Xiaobo Zhang\inst{3}* \and Rui Feng\inst{1,2,3}*}
\authorrunning{J. Hou et al.}
\institute{School of Computer Science, Shanghai Key Laboratory of Intelligent Information Processing, Fudan University, China\\
\and
Academy for Engineering and Technology, Fudan University, China\\
\email{\{jlhou18,jilanxu18,20210860062,yjzhang,fengrui\}@fudan.edu.cn}
\and Children’s Hospital of Fudan University, National Children’s Medical Center, Shanghai, China\\
\email{zhangxiaobo0307@163.com}\\}

% \end{comment}
%******************
\maketitle

\begin{abstract}
This paper presents our solution for the 2nd COVID-19 Severity Detection Competition. This task aims to distinguish the Mild, Moderate, Severe, and Critical grades in COVID-19 chest CT images. In our approach, we devise a novel infection-aware 3D Contrastive Mixup Classification network for severity grading. Specifically, we train two segmentation networks to first extract the lung region and then the inner lesion region. The lesion segmentation mask serves as complementary information for the original CT slices. To relieve the issue of imbalanced data distribution, we further improve the advanced Contrastive Mixup Classification network by weighted cross-entropy loss. On the COVID-19 severity detection leaderboard, our approach won the first place with a Macro F1 Score of 51.76\%. It significantly outperforms the baseline method by over 11.46\%.

\end{abstract}

\section{Introduction} 
\label{section:sec1}
The Coronavirus Disease 2019 SARS-CoV-2 (COVID-19) is an ongoing pandemic, which spreads fast worldwide since the end of 2019 \cite{WHO}. To date (July, 2022), the number of infected people is still increasing rapidly. In clinical practice, thoracic computed tomography (CT) has been recognized to be a reliable complement to RT-PCR assay \cite{fang2020sensitivity}. From CT scans, a detailed overview of COVID-19 radiological patterns can be clearly presented, including ground glass opacities, rounded opacities, enlarged intra-infiltrate vessels, and later more consolidations \cite{chung2020ct}. In fighting against COVID-19, two important clinical tasks can be conducted in chest CT scans, namely identification and severity assessment. Early identification is vital to the slowdown of viral transmission, whereas severity assessment is significant for clinical treatment planning.

Recently, deep learning approaches have achieved excellent performance in discriminating COVID-19 from non-pneumonia or other types of pneumonia \cite{Chen2020.02.25.20021568,javaheri2020covidctnet,HOU2021108005,wang2020a}. 
For instance, Chen et al. \cite{Chen2020.02.25.20021568} detected the suspicious lesions for COVID-19 diagnosis using a UNet++ model. Javaheri et al.\cite{javaheri2020covidctnet} designed the CovidCTNet to differentiate COVID-19 from other lung diseases. Xu et al. \cite{xuh1n1} developed a deep learning system with a location-attention mechanism to categorize COVID-19, Influenza-A viral pneumonia, and healthy subjects. Despite the great success in COVID-19 identification, the automatic assessment of COVID-19 severity in CT images is still a very challenging task.

\begin{figure}[t]
\centering
\includegraphics[width=0.85\textwidth]{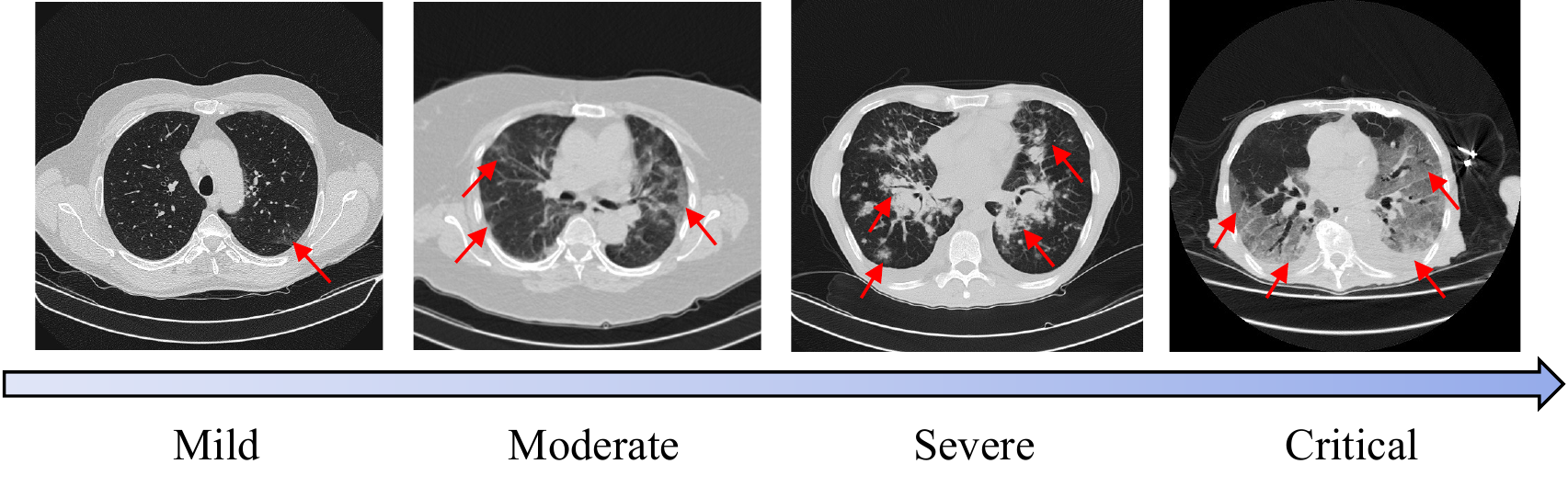}
\caption{Examples of COVID-19 CT images with different severity levels from the COV19-CT-DB dataset. The red arrows indicate the locations of the main lesion regions. In the mild case, the lesions only account for small regions in the lung. In the critical case, the lesions spread over the periphery of the lung and are indistinguishable from the outer tissue.}
\label{fig:intro}
\end{figure}

In this paper, we present a deep-learning based solution for the 2nd COV19D Competition of the Workshop ``AI-enabled Medical Image Analysis – Digital Pathology \& Radiology/COVID19 (AIMIA)'', which occurs in conjunction with the European Conference on Computer Vision (ECCV) 2022. This paper mainly focuses on the COVID-19 Severity Detection Challenge. As illustrated in Fig. \ref{fig:intro}, the goal of this task is to categorize the severity of COVID-19 into four stages, including Mild, Moderate, Severe, and Critical. Each severity category is determined by the presence of ground glass opacities and the pulmonary parenchymal involvement. This task is of great significance in real-world scenarios. Especially for patients with severe or critical grades, extensive care and attention are necessary in the clinical practice.
The main challenge of the severity detection task is three-fold: (1) Compared with the binary classification task (COVID vs. non-COVID), the severity detection requires more fine-grained lesion features to determine the final grade. (2) The accurate annotation of grades requires domain knowledge and it is labor-expensive. The limited amount of data makes the task difficult for deep-learning based methods. (3) In the original CT volume, it is hard to identify the lesions of interest from substantial non-lesion voxels. 

In order to address the aforementioned challenges, we propose a novel infection-aware Contrastive Mixup Classification network. This framework is constructed in a segmentation-classification manner.
In particular, we initially extract the lung regions and segment the inner lesion regions using two individual segmentation networks. The obtained lesion segmentation mask serves as complementary information for fine-grained classification. The location, as well as the number of lesions, are both crucial in determining the severity grade. After lesion segmentation, an advanced 3D Contrastive Mixup Classification network (CMC-COV19D) in the previous work \cite{hou2021cmc} is applied. It won the first price in the ICCV 2021 COVID-19 Diagnosis Competition of AI-enabled Medical Image Analysis Workshop \cite{kollias2021mia}. The CMC-COV19D framework introduces contrastive representation learning to discover more discriminative representations of COVID-19 cases. It includes a joint training loss that combines the classification loss, mixup loss, and contrastive loss. We further improve the classification loss with weight balancing strategy to enhance the attention on minor classes (i.e. moderate and critical). Experimental results on the COVID-19 severity detection challenge shows that our approach achieves a Marco F1 Score of 51.76\%, surpassing all the participants. 

The remainder of this paper is organized as follows. In Section \ref{section:relatedwork}, we summarize related works on segmentation of lung regions and lesions, and COVID-19 severity assessment. Section \ref{section:method} introduces our proposed infection-aware Contrastive Mixup Classification framework in detail. Section \ref{section:dataset} describes the two datasets. Extensive experiments
are conducted in Section \ref{section:experiments} to evaluate the performance of our network. The conclusion is given in Section \ref{section:conclusion}.

\section{Related Work} \label{section:relatedwork}

\subsection{Segmentation of lung regions and lesions}
Segmentation is an important step in AI-based COVID-19 image processing and analysis \cite{shi2020review}. It extracts the regions of interest (ROI), including lung, lobes, and infected regions in CT scans for further diagnosis. A large number of researches \cite{zheng2020deep,cao2020longitudinal,huang2020serial} utilized U-Net \cite{ronneberger2015u} to segment lung fields and lesions. For example, Qi et al. \cite{Qi2020.02.29.20029603} used U-Net to delineate the lesions in the lung and extract radiometric features of COVID-19 patients with the initial seeds given by a radiologist for predicting hospital stay. Moreover, Jin et al. \cite{tsinghua2020fourweek} proposed an COVID-19 screening system, in which the lung regions are first detected by UNet++ \cite{zhou2018unet++}. The Inf-Net \cite{fan2020inf} aimed to segment the COVID-19 infection regions for quantifying and evaluating the disease progression.
Shan et al. \cite{shan2020lung} proposed a VB-Net for segmentation of lung, lung lobes and lung infection, which provided accurate quantification data for medical studies. Motivated by the above approaches, we train segmentation networks to extract infection-aware clues for severity grading.

\subsection{COVID-19 severity assessment}
The study of COVID-19 severity assessment plays an essential role in clinical treatment planning. 
Tang et al. \cite{tang2020severity} established a random forest model based on quantitative features for COVID-19 severity assessment (non-severe or severe). Chaganti et al. \cite{chaganti2020automated} presented a method that automatically segmented and quantified abnormal CT patterns commonly presented in COVID-19.
Huang et al. \cite{huang2020serial} assessed a quantitative CT image parameter defined as the percentage of lung opacification (QCT-PLO) and classified the severity into four classes (mild, moderate, severe and critical).
Pu et al. \cite{pu2021automated} developed an automated system to detect COVID-19, quantify the extent of disease, and assess the progression of the disease.
Feng et al. \cite{feng2021severity} designed a novel Lesion Encoder framework to detect lesions in chest CT scans and encode lesion features for automatic severity assessment. 
He et al. \cite{he2021synergistic} proposed a synergistic learning framework for automated severity assessment of COVID-19 in 3D CT images. 
% It jointly performed lung lobe segmentation and multi-instance classification, where the segmentation task provides context information to aid the task of severity assessment in chest CT image. 
In our work, we first identify the infected regions, and then utilize segmented lesion masks as complementary information for the original CT slices.

\section{Methodology} \label{section:method}

\begin{figure}[t]
\centering
\includegraphics[width=\textwidth]{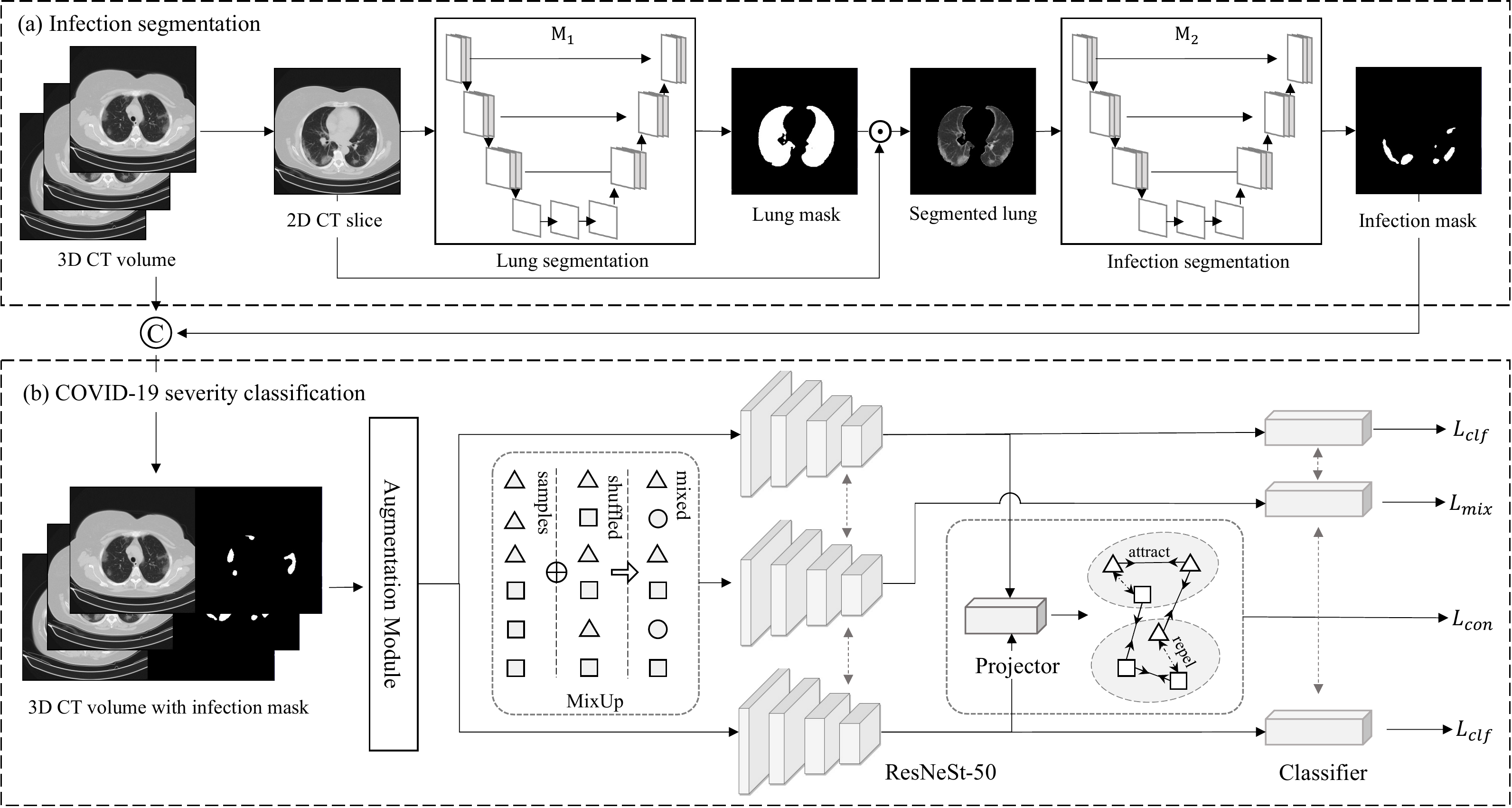}
\caption{Overview of our infection-aware CMC network, comprised of the infection segmentation stage and the COVID-19 severity classification stage. }
\label{fig:cmc}
\end{figure}

Fig. \ref{fig:cmc} illustrates the overview of our infection-aware Contrastive Mixup Classification (CMC) framework in a combined segmentation-classification manner. The framework is divided into two stages. The first stage is the infection segmentation that generates lesion masks of a CT volume slice by slice (Fig. \ref{fig:cmc}(a)).
The lesion masks of an input CT volume help the classification model focus on important regions and better classify COVID-19 severity. The second stage is to take a CT volume and its 3D lesion masks as the input and output the COVID-19 severity prediction (Fig. \ref{fig:cmc}(b)). We adopt our previous work, the CMC network, as the classification model, which has shown excellent COVID-19 detection performance.

In the following subsections, we will elaborate the infection segmentation procedure in detail, and then describe the CMC network for COVID-19 severity classification.

\subsection{Infection segmentation}
As shown in Fig. \ref{fig:cmc}(a), given an input CT volume $x=\{\mathbbm{x}_t\}_{t=1}^{T}$, we perform slice-level infection segmentation by the following two steps. 

\textbf{Step 1: lung region extraction.} For each slice $\mathbbm{x}_{t}\in\mathbb{R}^{H\times W}$ in the CT volume, we train a lung segmentation model $M_1$ to predict the lung region mask $M_1(\mathbbm{x}_t;\theta)\in\mathbb{R}^{H\times W}$, where $\theta$ are the network parameters. Extracting the lung regions of the CT volume removes the redundant background region as the background varies across different patients and screening devices. Moreover, lung extraction eases the subsequent lesion segmentation model as the lesions usually appear at the periphery of the lung. We empirically show that with simple training recipes, the lung segmentation model yields high segmentation accuracy. After training, the segmented CT slice $\hat{\mathbbm{x}_t}\in\mathbb{R}^{H\times W}$ is obtained by multiplying the lung segmentation mask with the original CT slice:
\begin{equation}
    \hat{\mathbbm{x}_t} = M_1(\mathbbm{x}_t;\theta) \odot \mathbbm{x}_{t}
\end{equation}
where $\odot$ denotes the Hadamard product. 

\textbf{Step 2: lesion segmentation.} Given each CT slice $\hat{\mathbbm{x}_t}$, we obtain the infection mask $I_t = M_2(\hat{\mathbbm{x}_t};\phi)$ by training another segmentation model $M_2$ parameterized by $\phi$. We refer the infection mask to the binary map where 1 indicates the pixel with COVID-19 infection and 0 represents the non-infected region.  
Notably, $M_2$ is trained on the CT slices with the segmented lung region, and thus the noisy background and artifacts are avoided. This reduces the model's potential false positive predictions over the background. The input to the classification model is defined as the concatenation of the infection mask and the original CT image:
\begin{equation}
    \bar{x}_t = [I_t, \mathbbm{x}_t] = [M_2(M_1(\mathbbm{x}_t;\theta) \odot \mathbbm{x}_{t};\phi), \mathbbm{x}_t].
\end{equation}

\subsubsection{Segmentation models.} 
We employ several state-of-the-art models for lung and infection segmentation, including DeepLabV3 \cite{chen2018encoder}, FPN \cite{lin2017feature}, U-Net \cite{ronneberger2015u}, and U-Net++ \cite{zhou2018unet++}. Different backbones such as ResNet \cite{he2016resnet} and DenseNet \cite{huang2017densely} with ImageNet pre-trained weights are adopted as the encoder layers of segmentation models. All segmentation models are trained and validated on the COVID19-CT-Seg dataset \cite{MP-COVID-19-SegBenchmark}. The models that perform best on the validation set are then transferred to the COV19-CT-DB dataset to generate lung and infection masks for each CT volume.

\subsubsection{Loss function.}
We use the Dice loss \cite{milletari2016v} as the objective function to optimize lung and infection segmentation networks. For each category $c$, the Dice loss is defined as follows:
\begin{equation}
    \mathcal{L}_{Dice}^c = 1- 2\times \frac{\sum_{n=1}^N y_n^{c} \hat{y}_n^c}{\sum_{n=1}^N (y_n^c+\hat{y}_n^c)+\epsilon},
\end{equation}
where $y_n$ and $\hat{y_n}$ denote the label and the prediction of each pixel $n$, respectively. $\epsilon$ is a small constant to avoid zero division.

\subsection{Infection-aware COVID-19 severity classification}

As illustrated in Fig. \ref{fig:cmc}(b), the CT volume $X\in\mathbb{R}^{T\times H\times W}$ and its infection mask $I\in\mathbb{R}^{T\times H\times W}$ are concatenated per channel to produce $\bar{X}\in\mathbb{R}^{2\times T\times H\times W}$ as the input. We adopt the Contrastive Mixup Classification (CMC) network for COVID-19 severity detection.

The CMC network is mainly comprised of the following four components.
\begin{itemize}
    \item \textbf{A data augmentation module $A(\cdot)$.} This module transforms an input CT volume $x$ into an augmented volume $\tilde{x}$ randomly. In our work, two augmented volumes are generated from each input CT scan.
    \item \textbf{A feature extractor $E(\cdot)$.} It maps the augmented CT sample $\tilde{x}$ to a representation vector $r=E(\tilde{x})\in \mathbb{R}^{d_e}$ in $d_e$-dimensional latent space. We adopt an inflated 3D ResNeSt-50 to extract features of each CT scan.
    \item \textbf{A small projection network $P(\cdot)$.} This network is used to map the high-dimension vector $r$ to a relative low-dimension vector $z=P(r)\in \mathbb{R}^{d_p}$. A MLP (multi-layer perception) can be employed as the projection network.
    \item \textbf{A classifier network $C(\cdot)$.} It classifies the vector $r\in \mathbb{R}^{d_e}$ to the COVID-19 severity prediction using fully connected layers. 
\end{itemize}

We train the CMC network using a well-designed adaptive loss, composed of contrastive loss, mixup loss, and weighted classification loss. 

\subsubsection{Contrastive loss.}
The CMC network employs the contrastive learning as an auxiliary task to learn more discriminative representations of each COVID-19 severity categories. Formally, given a minibatch of $2N$ augmented CT volumes and the labels $\{(\tilde{x}_i,\tilde{y}_i)\}_{i=1,\dots,2N}$, we define the positives as any augmented CT samples from the same category, while those from different classes are considered as negative pairs \cite{2020Supervisedcon}. Let $i\in \{1,\dots,2N\}$ be the index of an arbitrary augmented sample, the contrastive loss function is defined as:
\begin{equation}
    \mathcal{L}_{con}^i=\frac{-1}{2N_{\tilde{y}_i}-1}\sum_{j=1}^{2N}\mathbbm{1}_{i\ne j}\cdot\mathbbm{1}_{\tilde{y}_i=\tilde{y}_j}\cdot\log\frac{\exp(z_i^T\cdot z_j/\tau)}{\sum_{k=1}^{2N}\mathbbm{1}_{i\ne k}\cdot\exp(z_i^T\cdot z_k/\tau)},
\label{infonce}
\end{equation}
where $N_{\tilde{y}_i}$ is the number of samples in a minibatch that share the same label $\tilde{y}_i$, $\mathbbm{1}\in\{0,1\}$ is an indicator function, and $\tau$ denotes a temperature parameter.

\subsubsection{Mixup loss.}

To boost the generalization ability of the model, the mixup strategy \cite{zhang2017mixup} is also adopted in the training procedure. For each augmented CT volume $\tilde{x}_i$, we generate a mixed sample $\tilde{x}^{mix}_i$ and its mixed label $\tilde{y}^{mix}_i$ as:
\begin{equation}
    % \begin{aligned}
    \tilde{x}^{mix}_i = \lambda\tilde{x}_i + (1-\lambda)\tilde{x}_{p}, 
    ~\tilde{y}^{mix}_i = \lambda\tilde{y}_i + (1-\lambda)\tilde{y}_{p}, 
\end{equation}
where $p$ is a random index. The mixup loss is defined as the cross-entropy loss of mixed samples:
\begin{equation}
    \mathcal{L}^i_{mix} = \mathrm{CrossEntropy}(\tilde{x}^{mix}_i,\tilde{y}^{mix}_i).
\end{equation}

\subsubsection{Weighted classification loss.}
In addition to the mixup loss on the mixed samples, we also employ the cross-entropy loss on raw samples for severity detection. It is observed from Table \ref{table: dataset} that there are imbalance data distributions on the COV19-CT-DB dataset. To alleviate this problem, we adopt the weighted cross entropy loss. Each sample is scaled by the weight proportional to the inverse of the percentage of the sample for each class $c$ in the training set, denoted as $\alpha=[\alpha^1,\dots, \alpha^c]$.
Therefore, the weighted classification loss is defined as:
\begin{equation}
    \mathcal{L}_{clf}^i=-\alpha\tilde{y}_i^T\log\hat{y}_i,
\end{equation}
where $\tilde{y}_i$ denotes the one-hot vector of ground truth label, and $\hat{y}_i$ is predicted probability of the sample $x_i$ $(i=1,\dots,2N)$.

\subsubsection{Adaptive joint loss.}
Finally, we merge the contrastive loss, mixup loss, and weighted classification loss into a combined objective function with learnable weights \cite{kendall2018multi}:
\begin{equation}
     \mathcal{L}=\frac{1}{2N\sigma_1^2}\sum_{i=1}^{2N}\mathcal{L}^i_{con}+
     \frac{1}{2N\sigma_2^2}\sum_{i=1}^{2N}(\mathcal{L}^i_{mix}+\mathcal{L}^i_{clf})+\log\sigma_1+\log\sigma_2,
\end{equation}
where $\sigma_1, \sigma_2$ are utilized to learn the adaptive weights of the three losses.

% \subsection{(2+1)D-based Network}

\section{Dataset} \label{section:dataset}

\subsection{COV19-CT-DB Database}

The COV19-CT-Database (COV19-CT-DB) \cite{kollias2022ai} provides chest CT scans that are marked for the existence of COVID-19. It contains about 1,650 COVID and 6,100 non-COVID chest CT scan series from over 1,150 patients and 2,600 subjects. 
The CT scans were collected in the period from September 1, 2020 to November 30, 2021. 
Each CT scan was annotated by 4 experienced medical experts, which showed a high degree of agreement (around 98\%). The number of slices of each 3D CT scan ranging from 50 to 700. The variation in the number of slices is due to the context of CT scanning. 

In particular, the COV19-CT-DB further provides the severity annotations on a small portion of the COVID-19 cases. The descriptions and number of samples of the four severity categories are shown in detail in Table \ref{table: dataset}. The severity categories include Mild, Moderate, Severe, and Critical, in the range from 1 to 4. 
The training set contains, in total, 258 3D CT scans (85 Mild cases, 62 Moderate cases, 85 Severe cases, and 26 Critical cases). The validation set consists of 61 3D CT scans (22 Mild cases, 10 Moderate cases, 22 Severe cases, and 5 Critical cases). The testing set includes 265 scans and the labels are not available during the challenge.

\setlength{\tabcolsep}{4pt}
\begin{table}[t]
\begin{center}
\caption{The number of samples and descriptions of four severity categories on the COV19-CT-DB database.}
\label{table: dataset}
\resizebox{\linewidth}{!}{
\begin{tabular}{ccccc}
\hline\noalign{\smallskip}
Category & Severity & Training & Validation & Description \\
\noalign{\smallskip}
\hline
\noalign{\smallskip}
1& Mild & 85 & 22& \makecell{Few or no ground glass opacities. Pulmonary \\parenchymal involvement $\leq$ 25\% or absence}\\
\noalign{\smallskip}
\hline
\noalign{\smallskip}
2& Moderate & 62 & 10 & \makecell{Ground glass opacities. Pulmonary \\parenchymal involvement 25-50\%}\\
\noalign{\smallskip}
\hline
\noalign{\smallskip}
3& Severe & 85 & 22 & \makecell{Ground glass opacities. Pulmonary \\parenchymal involvement 50-75\%}\\
\noalign{\smallskip}
\hline
\noalign{\smallskip}
4& Critical & 26 & 5 & \makecell{Ground glass opacities. Pulmonary \\parenchymal involvement $\ge$ 75\%}\\
\noalign{\smallskip}
\hline
\end{tabular}
}
\end{center}
\end{table}
\setlength{\tabcolsep}{1.4pt}

\subsection{COVID-19-CT-Seg dataset}
The COVID-19-CT-Seg dataset \cite{MP-COVID-19-SegBenchmark} provides 20 public COVID-19 CT scans from the Coronacases Initiative and Radiopaedia. All the cases contain COVID-19 infections with the proportion ranging from 0.01\% to 59\%. The left lung, right lung, and infections (i.e., all visibly affected regions of the lungs) were annotated, refined and verified by junior annotators, radiologists, and a senior radiologist, progressively. The whole lung mask includes both normal and pathological regions. In total, there are more than 300 infections with over 1,800 slices. 
% The COVID-19-CT-Seg dataset are publicly available at \url{https://zenodo.org/record/3757476}.

\section{Experimental Results} \label{section:experiments}

\subsection{Implementation details}

\subsubsection{Lung and infection segmentation.}
We employ multiple segmentation models \cite{chen2018encoder,lin2017feature,ronneberger2015u,zhou2018unet++} with the ImageNet pre-trained encoder \cite{he2016resnet,huang2017densely} to perform lung and infection segmentation on 2D CT slices. The input resolution is resized to $256 \times 256$. Random crop, random horizontal flip, vertical flip, and random rotation are applied as forms of data augmentation to reduce overfitting.  
The initial learning rate is set to 1e-3 and then divided by 10 at $30\%$ and $80\%$ of the total number of training epochs. The segmentation models are trained for 100 epochs with the Stochastic Gradient Descent (SGD) optimizer. The batch size is 8. We select two models that perform best on the validation set as the final lung and infection segmentation models.

\subsubsection{COVID-19 severity detection.}
We employ an inflated 3D ResNest50 as the backbone for COVID-19 severity detection. Each CT volume is resampled from $T \times 512 \times 512$ to $64 \times 256 \times 256$, where $T$ denotes the number of slices. Data augmentations include random resized crop and random contrast changes. Other augmentations such as flip and rotation are also tried, but no significant improvement is yielded. We optimize the networks using the Adam algorithm with a weight decay of 1e-5. The initial learning rate is set to 1e-4 and then divided by 10 at $30\%$ and $80\%$ of the total number of training epochs. Our models are trained for 100 epochs with the batch size 4.
Our methods are implemented in PyTorch and run on four NVIDIA Tesla V100 GPUs. 

\subsection{Evaluation metrics}

For the lung and infection segmentation, we report mean Intersection over Union (mIoU), and IoU of the foreground. 
For the COVID-19 severity detection, we adopt the Macro F1 Score as the evaluation metric for overall comparison. The score is defined as the unweighted average of the class-wise/label-wise F1 Scores. We also show the F1 Score for each severity category. Furthermore, we utilize a model explanation technique, namely Class Activation Mapping (CAM) \cite{zhou2016learning}, to conduct visual interpretability analysis \cite{ioannou2021visual}. The CAM can identify the important regions that are most relevant to the predictions.

\subsection{Results of lung and infection segmentation}

% \setlength{\tabcolsep}{4pt}
% \begin{table}[t]
% \begin{center}
% \caption{The segmentation results on the COVID-19-CT-Seg dataset (unit: \%).}
% \label{table: segmentation}
% \resizebox{\linewidth}{!}{
% \begin{tabular}{lccccc}
% \hline\noalign{\smallskip}
% \multirow{2}{*}{Models} & \multirow{2}{*}{Encoder} & \multicolumn{2}{c}{Lung segmentation}&
% \multicolumn{2}{c}{Infection segmentation}\\
% \noalign{\smallskip}
% \cline{3-4} \cline{5-6}
% \noalign{\smallskip}
%  & & mAcc & mIoU (fg/bg) & mAcc & mIoU (fg/bg)\\

% \noalign{\smallskip}
% \hline
% \noalign{\smallskip}
% Deeplabv3 & Res34 & 98.41 & 95.01 (91.44/98.57) & 88.15 & 78.67 (58.10/99.25)\\
% FPN & Res34 & 98.59 & 96.81 (94.50/99.12) & 83.27 & 77.88 (56.46/99.30)\\
% UNet & Res34 & 98.55 & 96.96 (94.75/99.16) & 83.10 & 79.67 (59.93/99.40)\\
% UNet++ & Res34 & 98.86 & 97.37 (95.46/99.27) & 87.04 & 80.36 (61.36/99.36)\\
% UNet & Dense121 & 98.89 & 97.54 (95.75/99.32) & running\\
% UNet++ & Dense121 & 98.98 & 97.56 (95.79/99.33)\\
% \hline
% \end{tabular}
% }
% \end{center}
% \end{table}
% \setlength{\tabcolsep}{1.4pt}

\setlength{\tabcolsep}{4pt}
\begin{table}[t]
\begin{center}
\caption{The segmentation results on the COVID-19-CT-Seg dataset (unit: \%).}
\label{table: segmentation}
% \resizebox{\linewidth}{!}{
\begin{tabular}{lccccc}
\hline\noalign{\smallskip}
\multirow{2}{*}{Model} & \multirow{2}{*}{Encoder} & \multicolumn{2}{c}{Lung segmentation}&
\multicolumn{2}{c}{Infection segmentation}\\
\noalign{\smallskip}
\cline{3-4} \cline{5-6}
\noalign{\smallskip}
 & & mIoU & lung IoU & mIoU & infection IoU\\
\noalign{\smallskip}
\hline
\noalign{\smallskip}
FPN & ResNet-34 & 96.81 & 94.50 & 77.88 & 56.46\\
Deeplabv3 & ResNet-34 & 95.01 & 91.44 & 78.67 & 58.10\\
UNet & ResNet-34 & 96.96 & 94.75 & 79.67 & 59.93\\
UNet & DenseNet-121  & 97.54 & 95.75 & 77.52 & 55.73\\
UNet++ & ResNet-34 & 97.37 & 95.46 & \textbf{80.36} & \textbf{61.36}\\
UNet++ & DenseNet-121 & \textbf{97.56} & \textbf{95.79} & 80.04 & 60.71\\
\hline
\end{tabular}
% }
\end{center}
\end{table}
\setlength{\tabcolsep}{1.4pt}

\begin{figure}[t]
\centering
\includegraphics[width=\textwidth]{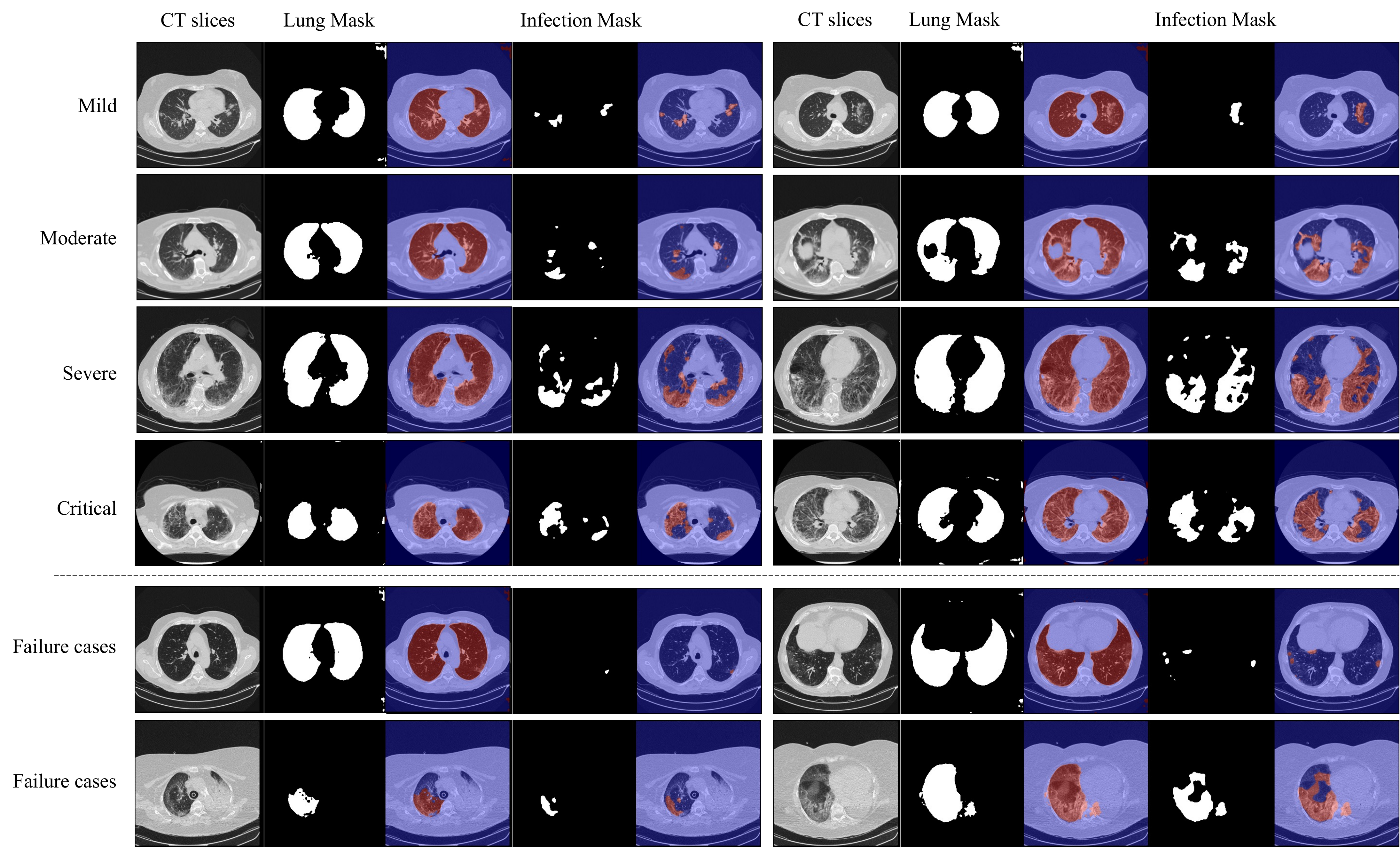}
\caption{An illustration of the segmentation results on the COV19-CT-DB dataset.}
\label{fig:segmentation}
\end{figure}

The lung and infection segmentation results on the COVID-19-CT-Seg dataset are shown in Table \ref{table: segmentation}. 
Among all segmentation models, the UNet++ models achieve better performance on both lung and infection segmentation tasks. The UNet++ with ResNet-34 encoder reaches the highest infection segmentation results with 80.36\% mean IoU and 61.36\% IoU for infections. The Unet++ with DenseNet-121 encoder performs best on the lung segmentation task. It obtains 97.56\% mIoU and 95.79\% IoU for lung regions. 

After the segmentation on the COVID-19-CT-Seg dataset, we utilize the two trained models to inference the lung and infection regions on the CT scans from COV19-CT-DB dataset. 
As the COV19-CT-DB dataset lacks the segmentation annotations, we illustrate several representative visual results in Fig. \ref{fig:segmentation}. For each severity category, the segmentation results of two slices from a 3D CT scan are presented in the first four rows. As can be apparently seen, our lung segmentation model can predict the whole lung regions accurately, which shows its strong generalization ability. In addition, our infection segmentation model can inference most infection areas from mild to critical precisely. 

We also analyze the failure cases, which are shown in the last two rows of Fig. \ref{fig:segmentation}. According to our observation, the incomplete segmentation results mainly occur in the following two cases. (1) In the early stage, the low-density infections with fuzzy boundaries are difficult to identify. (2) The appearance of consolidation may be similar to bones, which is also hard for the model to recognize. Therefore, it may be inaccurate to make severity predictions by directly calculating the volume ratio between the segmented infections and lung regions. Therefore, we make full use of the segmentation results to provide auxiliary information for the original CT scan inputs.

\subsection{Results on the COVID-19 severity detection challenge}

\setlength{\tabcolsep}{4pt}
\begin{table}[t]
\begin{center}
\caption{The comparison results on the validation set of COVID-19 severity detection challenge (unit: \%).}
\label{table: severity}
\resizebox{\linewidth}{!}{
\begin{tabular}{llccccccc}
\hline\noalign{\smallskip}
\multirow{2}{*}{ID} & \multirow{2}{*}{Methods} & \multirow{2}{*}{Con} & \multirow{2}{*}{Mixup} & \multirow{2}{*}{Macro F1} & \multicolumn{4}{c}{F1}\\
\cline{6-9}
\noalign{\smallskip}
&&&&& Mild & Moderate & Severe & Critical\\
\noalign{\smallskip}
\hline
\noalign{\smallskip}
1& CNN-RNN \cite{kollias2022ai}& -& - & 63.00 & -& -&-&-\\
2& ResNeSt-50 \cite{zhang2020resnest} &  &  &  65.47 & 82.05 & 61.54 & 68.29 & 50.00\\
3& ResNeSt-50 \cite{zhang2020resnest} & & $\checkmark$ &  67.30 & 71.79 & 75.00 & 60.87 & 61.54\\
4& CMC \cite{hou2021cmc} & $\checkmark$ & $\checkmark$ & 71.86 & 79.07 & 78.26 & 75.56 & 54.55\\
\noalign{\smallskip}
\hline
\noalign{\smallskip}
5 & Lung-aware CMC & $\checkmark$ & $\checkmark$ & 70.62 & 78.05  & 66.67  & 71.11 &  66.67\\
6 & Infection-aware CMC & $\checkmark$ & $\checkmark$ & 77.03 & 83.72 & 66.67 & 74.42 & 83.33\\
\hline
\end{tabular}
}
\end{center}
\end{table}
\setlength{\tabcolsep}{1.4pt}

\setlength{\tabcolsep}{4pt}
\begin{table}[t]
\begin{center}
\caption{The leaderboard on the COVID19 Severity Detection Challenge (unit: \%).}
\label{table:leaderboard}
\begin{tabular}{llcccccc}
\hline\noalign{\smallskip}
\multirow{2}{*}{Rank} & \multirow{2}{*}{Teams} & \multirow{2}{*}{Macro F1} & \multicolumn{4}{c}{F1}\\
\cline{4-7}
\noalign{\smallskip}
&&& Mild & Moderate & Severe & Critical\\
\noalign{\smallskip}
\hline
\noalign{\smallskip}
1& FDVTS (Ours) & 51.76 & 58.97 & 44.53 & 58.89 & 44.64\\
2& Code 1055 & 51.48  & 61.14  & 34.06 & 61.91 & 48.83 \\
3& CNR-IEMN & 47.11 & 55.67 & 37.88 & 55.46 & 39.46\\
4& MDAP & 46.00 & 58.14 & 42.29 & 43.78 & 40.03\\
5& etro & 44.92 & 57.69 & 29.57 & 62.85 & 29.57 \\
6& Jovision-Deepcam & 41.49 & 62.20 & 40.24 & 38.59 & 24.93\\
7& ResNet50-GRU \cite{kollias2022ai} & 40.30 & 61.10 & 39.47 & 37.07 & 23.54\\

\hline
\end{tabular}
\end{center}
\end{table}

\subsubsection{Results on the validation set.}
Table \ref{table: severity} shows the results of the baseline model and our methods on the validation set of COVID-19 severity detection challenge. 
The baseline model is based on a CNN-RNN architecture \cite{kollias2022ai} that performs 3D CT scan analysis, following the work \cite{kollias2020deep,kollias2020transparent,kollias2018deep} on developing deep neural architectures for predicting COVID-19. It achieves 63.00\% Marco F1 Score.
In our experiments, a simple 3D ResNeSt-50 model reaches better performance with 65.47\% Macro F1 compared to the baseline. The Mixup technique helps to boost the detection result by $\sim$2\% improvement. When we adopt the CMC architecture with Contrastive learning and Mixup, it further improves the detection performance with 71.86\% Macro F1 Score. In addition, the CMC model significantly increases the F1 Scores on Moderate and Severe categories, which are easily confused. The experimental results demonstrate the effectiveness of Contrastive learning and Mixup modules.

Furthermore, we report the performance of our proposed lung-aware CMC and infection-aware CMC in the 5th and 6th rows in Table \ref{table: severity}. It can be seen that the combination of original images and lung region masks does not significantly benefit the severity detection results. However, the infection-aware CMC which takes the concatenation of original images and infection region masks as the input significantly boost the performance to 77.03\% Macro F1 Score. Additionally, this model shows the best or comparable results on the Mild (83.72\%), Moderate (66.67\%), Severe (74.42\%), and Critical (83.33\%) categories.

\begin{figure}[h]
\centering
\subfigure{
\begin{minipage}[t]{\textwidth}
\centering
\includegraphics[width=\textwidth]{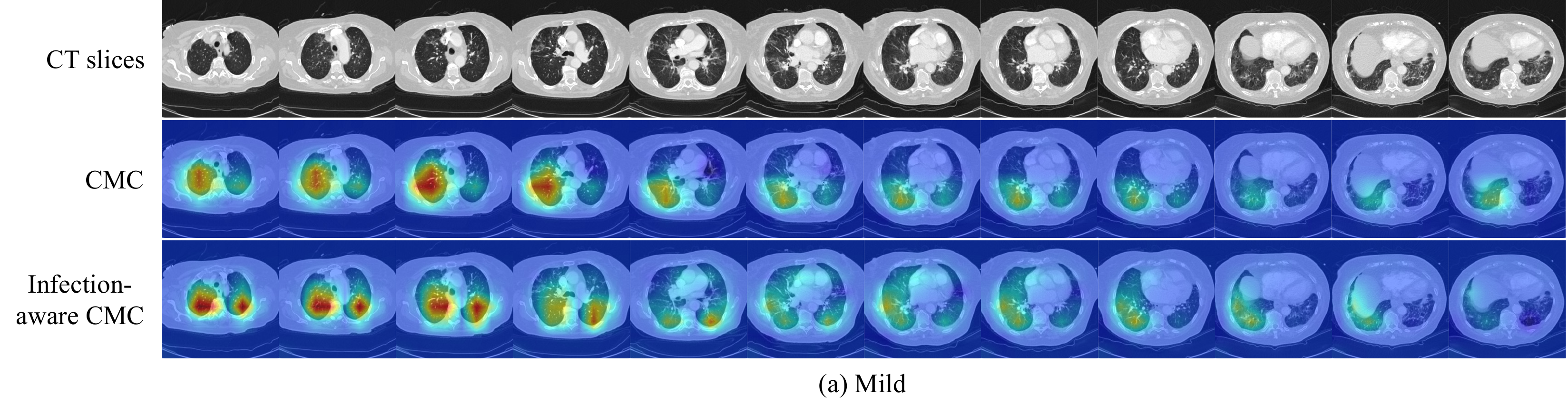}
%\caption{fig1}
\end{minipage}%
}%
\quad
\subfigure{
\begin{minipage}[t]{\textwidth}
\centering
\includegraphics[width=\textwidth]{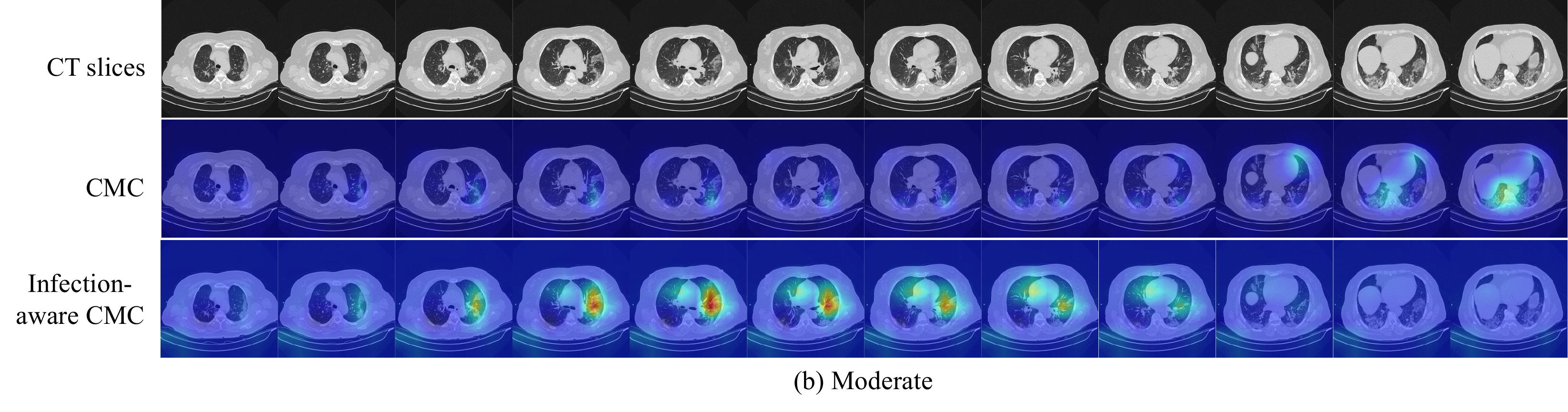}
%\caption{fig2}
\end{minipage}%
}%
\quad
\subfigure{
\begin{minipage}[t]{\textwidth}
\centering
\includegraphics[width=\textwidth]{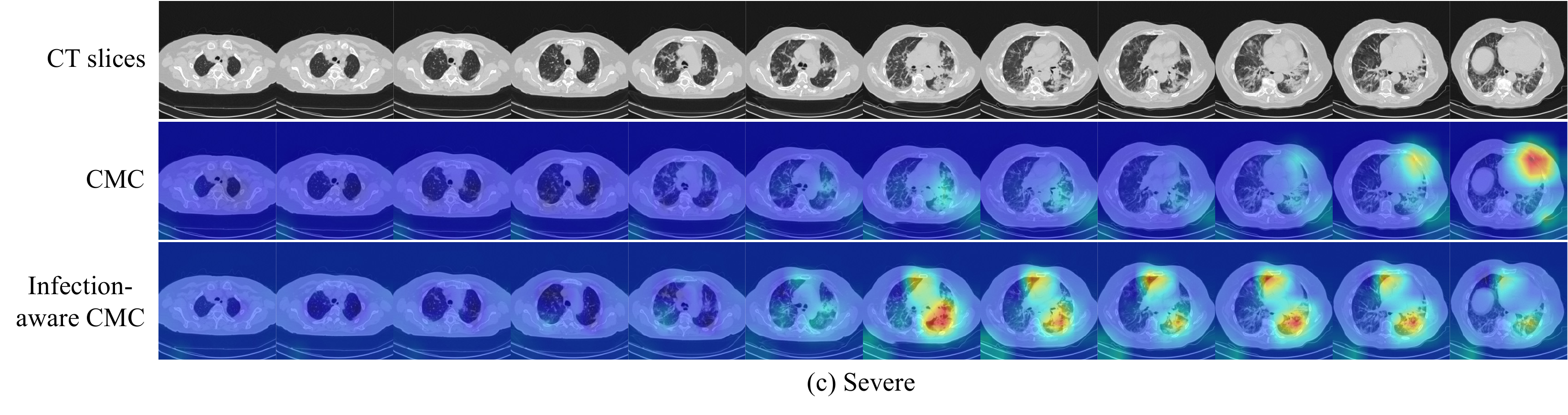}
%\caption{fig2}
\end{minipage}
}%
\quad
\subfigure{
\begin{minipage}[t]{\textwidth}
\centering
\includegraphics[width=\textwidth]{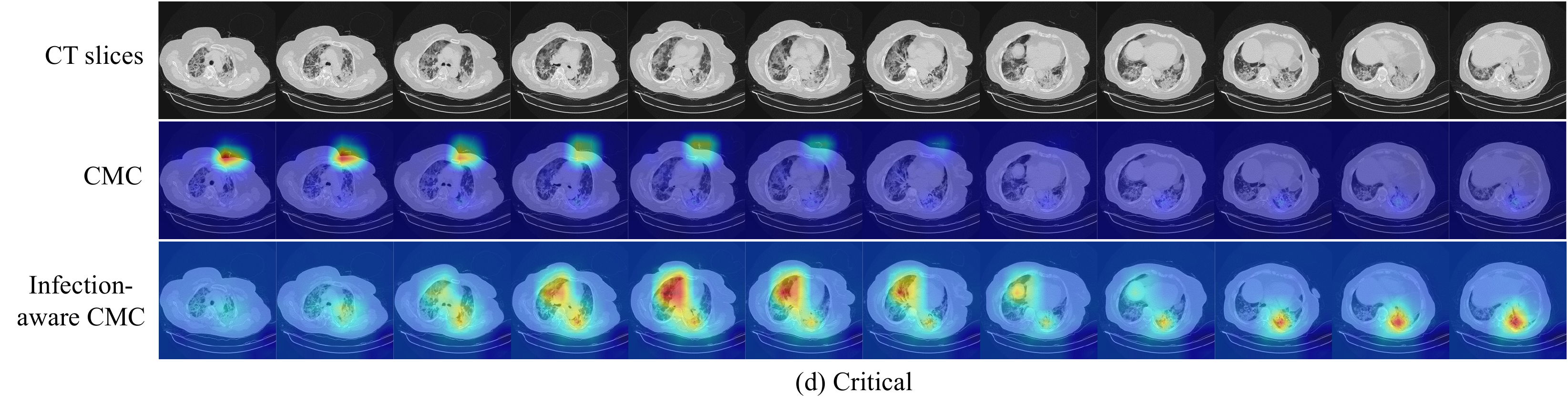}
%\caption{fig2}
\end{minipage}
}%
\centering
\caption{The visualization of model outputs on CT scan of different severity categories. }
\label{fig: cam}
\end{figure}

\subsubsection{The leaderboard on the COVID-19 severity detection challenge.}
In the COVID-19 severity detection challenge, we ensemble the CMC and infection-aware CMC models as our final model. The comparison results on the testing dataset are shown in Table \ref{table:leaderboard}. Our model ranks first in the challenge, surpassing the second team by 0.28\% on the Macro F1 Score. Compared to other methods, our model shows relatively stable performance on each category, indicating that our model has learned the discriminative representations of different severity categories.

\subsubsection{Visualization results.}
We exploit the Class Activation Mapping (CAM) \cite{zhou2016learning} to show the attention maps for COVID-19 severity detection. Four CT scan series of different categories are selected from the validation set, and the visualization results are shown in Figure \ref{fig: cam}. For each group, the first row shows a series of CT slices, and the second and third rows present the CAMs from CMC and infection-aware CMC models, respectively. 

Though the CMC model makes correct predictions, the attention maps just roughly indicate the suspicious area (see subfig. (a)), and it even over-focuses on irrelevant regions (e.g. bones) rather than the infections (see subfig. (b)(c)(d)). 
In contrast, our proposed infection-aware CMC model greatly improves the localization ability of suspicious lesions. Most infection areas can be located precisely, and the irrelevant regions are largely excluded. These highlighted areas in the attention maps demonstrate the reliability and interpretability of the detection results from our model.

\section{Conclusions} \label{section:conclusion}
In this paper, we present our solution for COVID-19 severity detection challenge in the 2nd COV19D Competition.
Our proposed infection-aware 3D CMC network is composed of two stages, namely infection segmentation and COVID-19 severity classification. In the first stage, two segmentation networks are trained to extract lung fields as well as segment infected regions. The infection masks further provide auxiliary information for the original CT scans.
In the second stage, the CMC network takes the CT scan and its lesion masks as the input and outputs the severity prediction. The experimental results on the competition leaderboard demonstrate the effectiveness of our infection-aware CMC network.

\section*{Acknowledgement}
This work was supported by the Scientific \& Technological Innovation 2030 - ``New Generation AI'' Key Project (No. 2021ZD0114001; No. 2021ZD0114000), and the Science and Technology Commission of Shanghai Municipality (No. 21511104502; No. 21511100500; No. 20DZ1100205). Yuejie Zhang, Xiaobo Zhang, and Rui Feng are corresponding authors.

\bibliographystyle{splncs04}
\bibliography{egbib}
\end{document}